\documentclass[11pt,a4paper]{article}
\usepackage{jheppub}
\usepackage{setspace,braket}
\usepackage{amsmath, amssymb, bm}
\usepackage{etoolbox}

\def\d#1#2{\frac{\displaystyle #1}{\displaystyle #2}}
\def\be{\begin{equation}}
\def\ee{\end{equation}}
\def\bea{\begin{eqnarray}}
\def\eea{\end{eqnarray}}
\def\no{\nonumber}
\def\pt{\partial}

\def\({\left(}
\def\){\right)}
\def\[{\left[}
\def\]{\right]}

\def\ra{\rightarrow}

\numberwithin{equation}{section}
\renewcommand{\thesection}{\arabic{section}}

\begin{document}
\renewcommand{\thefootnote}{\fnsymbol{footnote}}

\title{Fractional-order phase transition of charged AdS black holes}
\author[a,b]{Meng-Sen Ma}
\affiliation[a]{Institute of Theoretical Physics, Shanxi Datong
University, Datong 037009, China}
\affiliation[b]{Department of Physics, Shanxi Datong
University, Datong 037009, China}

\emailAdd{mengsenma@gmail.com}

\abstract{

Employing the fractional derivatives, we construct a more elaborate classification of phase transitions compared to the original Ehrenfest classification. In this way, a thermodynamic system can even undergo a fractional-order phase transition.
We use this method to restudy the charged AdS black hole and Van der Waals fluids and find that at the critical point they both have a $4/3$-order phase transition, but not the previously recognized second-order one.

}
\maketitle
\onehalfspace

\renewcommand{\thefootnote}{\arabic{footnote}}
\setcounter{footnote}{0}
\section{Introduction}
\label{intro}

In many thermodynamic systems there exist phase transitions and critical phenomena. The original classification of phase transitions was proposed by Ehrenfest. Specifically, when the $n$-th derivative of a thermodynamic potential (usually the Gibbs free energy or Helmholtz free energy) with respect to the external fields has a jump discontinuity while its lower order derivatives are all continuous, the thermodynamic system undergoes a $n$-order phase transition at the transition points. When $n\geq 2$, it is usually called the continuous phase transition. Fisher generalized the scope of continuous phase transitions by allowing for the divergent $n$-th derivative of thermodynamic potentials near the critical point\cite{Stanley.1987}.

Mathematically, there is also non-integer order derivatives, or fractional derivatives for short. This tool has been used in many areas of physics\cite{Riewe.1997,Laskin.2000,Sokolov.2001}. Following Ehrenfest's idea, one can employ the fractional derivative and generalize the classification of phase transition to include the fractional order. Therefore, the generalized Ehrenfest classification should have the following form:
\be
\lim_{T\ra T_c^{+}}\d{d^\alpha G}{dT^\alpha}=A^{+}\neq A^{-}=\lim_{T\ra T_c^{-}}\d{d^\alpha G}{dT^\alpha},
\ee
where $G=G(T,P)$ is the Gibbs free energy and $\alpha$ is a positive real number.
We expect that this finer-grained classification can differentiate phase transitions more effectively  compared to the relatively coarse-grained Ehrenfest classification.

The generalized classification can work in any thermodynamic systems that having phase structures. Black holes, being characterized by only three parameters, are the most simplest thermodynamic system. In particular, it has been found that AdS black holes have fruitful phase structures\cite{Hawking.1983,Chamblin.1999,Chamblin.1999b,Peca.1999,Wu.2000,Myung.2008,Quevedo.2008,Cadoni.2010,Liu.2010,Sahay.2010,Banerjee.2011,Ma.2014,Ma.2014b}. Therefore, in this paper we choose the charged AdS black hole to make the first attempt.

The plan of this paper is as follows:
In Sec.2 we briefly introduce the history and some useful formulae on fractional derivatives. 
In Sec.3 we analyze the fractional phase transition of the charged AdS black hole at the critical point. We also study the fractional phase transition of Van der Waals fluids for comparison.
In Sec.4 we summarize our results and discuss the possible future directions.(We will use the units: $\hbar=c=G=k_B=1$)

\section{Useful formulae on fractional derivatives}

In this part we only briefly introduce the history of fractional derivatives and give some formulae needed in the following sections. The strict definitions of fractional derivatives are left in Appendix \ref{fd}.

The idea of derivatives of fractional order has a long history, which can date back to the era of L'H\^{o}pital and Leibniz. After that, many mathematicians had considered the subject. The detailed history on this subject can be found in \cite{OS.1974}. The first documented example on fractional derivative was given by Lacroix (in 1819), who generalized the derivatives of power function by using Gamma function to replace the factorial of an integer:
\be\label{fdla}
\frac{d^mx^n}{dx^m}=\frac{n!}{(n-m)!}x^{n-m} \Rightarrow \frac{d^\mu x^\nu}{dx^\mu}=\frac{\Gamma(\nu+1)}{\Gamma(\nu-\mu+1)}x^{\nu-\mu},
\ee
where $(m,~n)$ are integers and $(\mu,~\nu)$ are real numbers. The Gamma function is defined as $\Gamma(\mu)\equiv \int_0^\infty x^{\mu-1}e^{-x}dx, ~~(\mu>0) $. The generalized formula can give us, for example, $\frac{d^{1/2}x}{dx^{1/2}}=\frac{2}{\sqrt{\pi}}x^{1/2}$ and especially $\frac{d^{1/2} 1}{dx^{1/2}}=\frac{1}{\sqrt{\pi}}x^{-1/2}$. Surprisingly, the fractional derivative of a constant returns a non-zero result. This is indeed the characteristic of the Riemann-Liouville fractional derivative.

In this paper, we prefer to choose the Caputo's definition of fractional derivatives, which will give us more acceptable results. Here we only list some formulae that will be used next.  We will write $D_x^{\alpha}=\frac{d^{\alpha} }{dx^{\alpha}}$ for simplicity.
First,
\be
D_x^{\alpha}c=0, ~~ (\alpha>0); \quad D_x^{\alpha}x=0, ~~(\alpha>1),
\ee
where $c$ is a constant and $\alpha$ is a real number. For $1<\alpha<2$, there are
\be\label{fdform}
D_x^{\alpha}x^2=\frac{2 x^{2-\alpha}}{\Gamma(3-\alpha)}, ~~~(x>0); \quad 
D_x^{\alpha}x^2=-\frac{2 (-x)^{2-\alpha}}{\Gamma(3-\alpha)}.~~~~(x<0).
\ee

\section{Charged AdS black holes}

If treating the cosmological constant as the thermodynamic pressure $P=-\Lambda/8\pi$, one can construct an extended phase space for AdS black holes\cite{Kastor.2009,Dolan.2011}. In the extended phase space, one can construct a one-to-one correspondence between the thermodynamic quantities of AdS black hole and those of ordinary thermodynamic systems\cite{Kubiznak.2012}. After that, many other AdS black holes and other interesting critical phenomena have been extensively explored\cite{Cai.2013,LYX.2013,Chen.2013,Hendi.2013,Altamirano.2014,Mo.2014,Xu.2014, Bin.2014, Ma.2015,Ma.2017,Sheykhi.2018}.

The equation of state for the charged AdS black hole can be written as
\be
P=\frac{T}{v}-\frac{1}{2\pi v^2}+\frac{2Q^2}{\pi v^4},
\ee
where the specific volume $v$ is related to the horizon radius, $v=2r_{+}$. The critical point lies at
\be
T_c=\d{\sqrt{6}}{18\pi Q}, \quad v_c=2\sqrt{6}Q, \quad P_c=\d{1}{96\pi Q^2}.
\ee

For simplicity, we define the dimensionless quantities,
\be\label{dimpvt}
p=\d{P-P_c}{P_c}, ~~~ \nu=\d{v-v_c}{v_c}, ~~~ t=\d{T-T_c}{T_c}.
\ee
With this new set of variables $({t}, {p}, {\nu})$, the critical point lies at $({t}={p}={\nu}=0)$.

Now the equation of state becomes
\be\label{eosbh}
(3 p+3)\nu ^4 + (12 p-8 t+4)\nu ^3+ (18 p-24 t)\nu ^2 + (12 p-24 t)\nu +3 p-8 t=0,
\ee
which is a quartic equation for $\nu$ and can be exactly solved in principle, although the roots are very complicated.

The dimensionless Gibbs free energy has the form
\be
g(t,p)=\frac{8-\nu ^4-4 \nu ^3+8 \nu -(\nu +1)^4 p}{4 \sqrt{6} (\nu +1)},
\ee
and the corresponding entropy is
\be
s(t,p)=-\left.\frac{\partial g}{\partial t}\right|_p=\sqrt{\frac{2}{3}}(1+\nu)^2,
\ee
where $\nu=\nu(p,t)$ is given by Eq.(\ref{eosbh}).

\begin{figure}
	\center{
		\includegraphics[width=7cm]{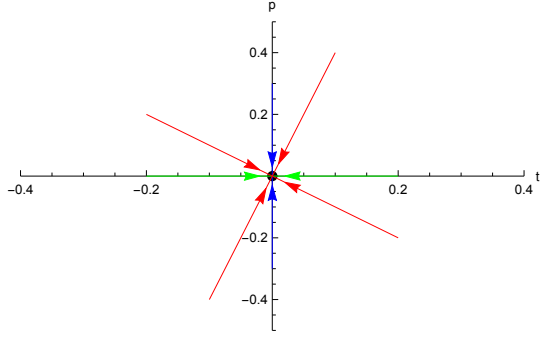}
		\caption{The black point at $(0,0)$ represents the critical point. The colorful arrows denote the various directions approaching the critical point.}\label{figcp}}
\end{figure}

We want to know the behaviors of the Gibbs free energy and its derivatives when the critical point is approached. Mathematically, for a function of two variables its continuity at certain point should be independent of the ways we choose to approach the point. As is shown in Fig.\ref{figcp}, there are many possible paths of approaching the critical point in the $({t}, {p})$ plane. It can be easily found that there are always $g(0,0)=2/\sqrt{6}$ and $s(0,0)=\sqrt{2/3}$, no matter which direction we choose to approach the critical point. Therefore, at the critical point the Gibbs free energy and the entropy are both continuous. So we only need to pay attention to the behavior of higher-order derivatives of the Gibbs free energy at the critical point.

Substituting $ \nu( t, p)$ derived from Eq.(\ref{eosbh}) into the Gibbs free energy and expanding it as a series of $ t$ and $ p$, we have
\bea\label{gRN}
g(t,p)&=&\left[\sqrt{\frac{2}{3}}+\frac{p}{2 \sqrt{6}}+...\right]-\left[\sqrt{\frac{2}{3}}-\frac{2^{5/6} p^{1/3}}{3^{1/6}}+...\right]t \no \\
&-&\left[\frac{4\times 2^{5/6}}{9 \times 3^{1/6} p^{2/3}}-\frac{26\times 2^{1/6}}{9\times 3^{5/6} p^{1/3}}+...\right]t^2 + ...
\eea
The type of series in the square bracket with fractional exponents is called a Puiseux series, we will see below, which is the key element for the fractional phase transition.

One can formally express the above result in the Taylor series of $ t$,
\be
g( t, p)=A( p)+B( p) t+D( p) t^2+O[ t^3].
\ee
The heat capacity at constant pressure $C_p$ is proportional to $\left.\frac{\pt^2 g}{\pt  t^2}\right|_{ p}$, so near the critical point the coefficient $D( p)$ is proportional to $C_p$. Generally, the coefficient $D( p)$ is divergent at the critical point, so we have the divergent $C_p$, which is a signal of the second-order phase transition according to the conventional Ehrenfest classification.

However, if taking into account the fractional derivative, the results can be completely different. According to Eq.(\ref{fdform}), the $\alpha$-order fractional derivatives of the Gibbs free energy with $1<\alpha\leq 2$ are
\bea\label{bhdg1}
\left.D_{ t}^{\alpha}g( t, p)\right|_{ p}&=&-\frac{4\times 2^{1/6} \left(2\times 6^{2/3}-13 p^{1/3}+...\right) }{9\times3^{5/6} \Gamma (3-\alpha )}\no \\
&\times& \frac{ t^{2-\alpha }}{ p^{2/3}}, ~~~~( t>0)
\eea
and
\bea\label{bhdg2}
\left.D_{ t}^{\alpha}g( t, p)\right|_{ p}&=&\frac{4\times 2^{1/6} \left(2\times 6^{2/3}-13 p^{1/3}+...\right) }{9\times3^{5/6} \Gamma (3-\alpha )}\no \\
&\times& \frac{(- t)^{2-\alpha }}{ p^{2/3}}, ~~~~( t<0).
\eea
Now we should calculate the values of $D_{ t}^{\alpha}g( t, p)$ in the limit $( t \rightarrow 0,~ p\rightarrow 0)$. As we mentioned above, mathematically there are infinitely ways of approaching the critical point. However, physically, due to the constraint of the equation of state, near the critical point $p$ and $t$ are not independent. Between them there is the relation\cite{KB.2017}
\be\label{vdwpt}
p =k t +O[ t^{3/2}],
\ee
with some constant $k$. Substituting it into Eq.(\ref{bhdg1}) and Eq.(\ref{bhdg2}), and take the limit, we can obtain
\be
\lim_{ t\rightarrow 0^{\pm}}D_{ t}^{\alpha}g( t, p)=\left\{
\begin{array}{lr}
	0 ~~&\text{for}~~\alpha<4/3,\\
	\mp\frac{4 \times 2^{5/6}}{3\times 3^{1/6}k^{2/3} \Gamma \left(\frac{2}{3}\right)} ~~&\text{for}~~\alpha=4/3,\\
	\mp\infty ~~ &\text{for}~~\alpha>4/3,
\end{array}
\right.
\ee

\begin{figure}
	\center{
		\includegraphics[width=8cm]{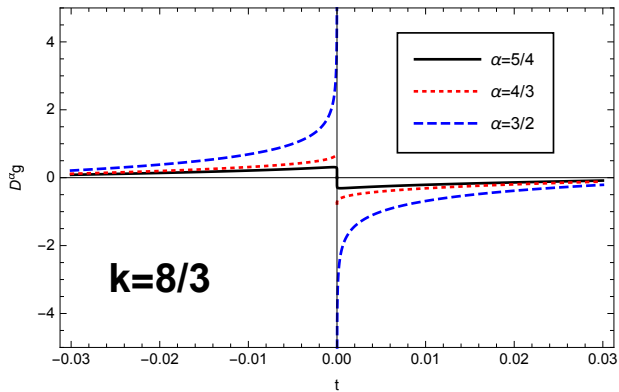}
		\caption{The behaviors of $D_{ t}^{\alpha}g$ near the critical point for the charged AdS black hole.}\label{figrndg}}
\end{figure}

Obviously, in the $\alpha=4/3$ case, there is
\be
\lim_{ t\rightarrow 0^{-}}D_{ t}^{\alpha}g \neq \lim_{ t\rightarrow 0^{+}}D_{ t}^{\alpha}g,
\ee
which is a jump discontinuity. When $\alpha > 4/3$, the $\alpha$-order fractional derivatives of the Gibbs free energy diverge. Therefore, the phase transition of the Van der Waals fluid at the critical point is of $4/3$ order according to the generalized Ehrenfest classification. In particular, the $k$ dependence of the results manifests that the discontinuity is universal along any directions except for $k=0$ and $k=\infty$. In fact, due to the constraint of the equation of state, $k$ cannot take arbitrary values.
From Eq.(\ref{eosbh}), one can easily see that
\be
p \approx \frac{8 t}{3 (\nu +1)},
\ee
near the critical point. Therefore, $k$ must be very close to $8/3$ for the charged AdS black hole\cite{Kubiznak.2012}.
In Fig.\ref{figrndg}, we show how $D_{ t}^{\alpha}g$ varies with $ t$. It can be found that the curves of $D_{ t}^{\alpha}g$ are symmetric about the critical point (the origin). The continuity, jump discontinuity and divergent behavior of $D_{ t}^{\alpha}g$ at different values of $\alpha$ are clearly presented.

For comparison, we also analyze the fractional phase transition of Van der Waals fluids. Van der Waals equation, which was proposed to describe the behavior of real fluids, has the form
\be
\left(P+\frac{a}{v^2}\right)(v-b)=T,
\ee
where $P$ is the thermodynamic pressure, $T$ is the temperature, and $v=V/N$ is the specific volume of the fluid.
The two parameters $a$ and $b$ are introduced to describe the interaction and size of the molecules in real fluids.
Van der Waals equation can be used to describe the critical phenomena of real fluids. The critical point lies at
\be
T_c=\d{8a}{27b},\quad v_c=3b,\quad P_c=\d{a}{27b^2}.
\ee
Below the critical point, Van der Waals equation can describe the first-order (liquid/gas) phase transition of fluids by taking account of the Maxwell's equal area law.

\begin{figure}
	\center{
		\includegraphics[width=8cm]{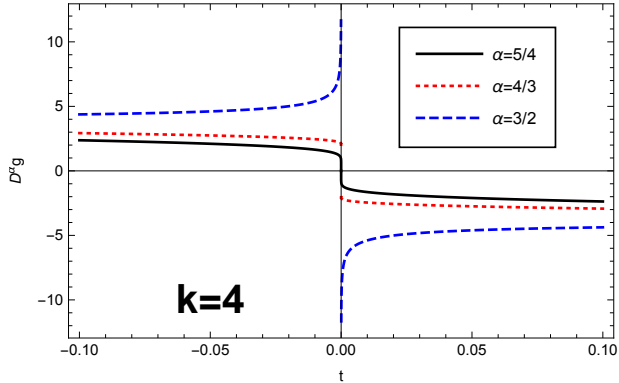}
		\caption{The behaviors of $D_{ t}^{\alpha}g$ near the critical point for Van der Waals fluids.}\label{figvdwg}}
\end{figure}

We also employ the dimensionless variables defined in Eq.(\ref{dimpvt}). Van der Waals equation turns into
\be\label{vdwpvt}
\nu ^3 (3  p+3)+\nu ^2 (8  p-8  t)+\nu  (7  p-16  t)+2  p-8  t=0.
\ee
The dimensionless Gibbs free energy and entropy are
\bea\label{vdwg0}
g( t, p)&=&-g_0 ( t+1)-\frac{3}{\nu +1}+(\nu +1) ( p+1)  \\
&-&\frac{8}{3} ( t+1) \ln \left(\nu +\frac{2}{3}\right)-4 ( t+1) \ln ( t+1),\no
\eea
and
\be\label{vdws}
s( t, p)=g_0+4+\frac{8}{3} \ln \left( \nu +\frac{2}{3}\right)+4 \ln ( t+1),
\ee
respectively, where $g_0$ is a constant.

Similarly, we expand $g(t,p)$ in series of $t$ and $p$,
\bea\label{gvdw}
g( t, p)&=&\left[-2-g_0+\frac{8}{3} \ln\frac{3}{2}+ p+...\right]+\left[-4-g_0+\frac{8}{3} \ln\frac{3}{2}+4\left(\frac{2}{3}\right)^{1/3} p^{1/3}+...\right] t \no \\
&-&\left[\frac{8}{3}\left(\frac{2}{3}\right)^{1/3}  p^{-2/3}+\frac{16}{9} \left(\frac{2}{3}\right)^{2/3} p^{-1/3}+...\right] t^2+ O[ t^3].
\eea
Clearly, Eq.(\ref{gvdw}) has nearly the same form as Eq.(\ref{gRN}).
For Van der Waals fluids, near the critical point there is still  $( p =k t)$ approximately. The only difference is that now $k$ should take values near $k=4$. In a similar way, we find that
\be
\lim_{ t\rightarrow 0^{\pm}}D_{ t}^{\alpha}g( t, p)=\left\{
\begin{array}{lr}
	0 ~~~~&\text{for}~~\alpha<4/3,\\
	\mp\frac{8 \left(\frac{2}{3}\right)^{1/3}}{k^{2/3} \Gamma \left(\frac{2}{3}\right)} ~~&\text{for}~~\alpha=4/3,\\
	\mp\infty ~~~ &\text{for}~~\alpha>4/3,
\end{array}
\right.
\ee
By the same token, the Van der Waals fluids should also undergo a $4/3$-order phase transition at the critical point. For comparison, we also depict the curves of $D_{ t}^{\alpha}g$ for different $\alpha$ in Fig.\ref{figvdwg}. It can be seen that the jump discontinuity behaviors of $D_{ t}^{\alpha}g$ for the charged AdS black holes and the Van der Waals fluids are very similar. The only difference is that the slope of the $\alpha=4/3$-curve for the charged AdS black hole is positive near the critical point, while the slope is negative for the Van der Waals fluids.

Several comments are in order at this stage:

(1).In this paper we only considered the fractional derivative with respect to $t$, namely $D_{ t}^{\alpha}g$. One can also discuss the $\alpha$-order fractional derivatives of the Gibbs free energy with respect to $p$. In that case one should first expand $g(t,p)$ as Taylor series of $p$, then expand the coefficients as series of $t$. You will find that the series about $t$ is the Puiseux series. Following the above calculation, one can draw the same conclusions.

(2).Fluid system and magnetic system are often compared to each other. In the mean field approximation, not only do they have similar thermodynamic relations, but also similar critical behaviors. Now we can differentiate their phase transitions. For magnetic system, due to the jump discontinuity of heat capacity at the critical point, it is the true second-order phase transition. While the Van der Waals fluid has $4/3$-order phase transition at the critical point.

(3). Hilfer also discussed the fractional phase transition and multiscaling issue\cite{Hilfer.2000,Hilfer:1992}. Starting from a thermodynamic potential, such as the Gibbs free energy $g(t,p)$ with the critical point at $(t=0,~p=0)$, Hilfer first introduced a curve $\mathcal{C}$ parameterized by $s$ in the phase space, $(t(s),p(s))$. Thus, at the beginning the two variables $(t,~p)$ are not independent. He then calculate the fractional derivatives
\be
A^{\pm}(\mathcal{C})=\lim_{s \rightarrow 0^{\pm}}\frac{d^\alpha g(t(s),p(s))}{ds^\alpha},
\ee
with real number $\alpha$. We completely follow the conventional thermodynamic definitions and take $(t,~p)$ as independent variables at the beginning. After deriving the fractional derivatives of the Gibbs free energy, $D_{ t}^{\alpha}g$, we compute their values in the limit $(t\ra 0, ~p\ra 0)$ and judge the phase transition correspondingly.

Our conclusions are different from, even opposite to those of Hilfer. For Van der Waals fluids, Hilfer's results indicate that the direction labelled by $k=4$ is special, along which the phase transition is of second order, while along other directions except for $k=4$ the phase transition is of $4/3$ order. We find that the order of the phase transition along the $k=4$ direction is $4/3$. What is more, near the critical point one can only discuss the phase transition along this direction. This issue was often overlooked in the previous researches on second-order phase transition, because the response functions, $c_p,~\kappa_T,~\alpha_T$ etc. are all divergent at the critical point, no matter from which directions the critical point is approached.

\section{Conclusion and Discussion}
\label{Conclusions}

With the aid of fractional derivatives,  one can calculate not only  the integer-order derivatives of thermodynamic potentials, but also any positive fractional-order derivatives of them. This allows us to generalized the original Ehrenfest classification of phase transition. Now we can classify phase transitions according to the jump discontinuity of fractional-order derivatives of thermodynamic potentials.  As examples, we reanalyzed the phase transition of the charged AdS black hole and the Van der Waals fluids at the critical point. we find that the $4/3$-order derivative of the Gibbs free is discontinuous at the critical point. What is more, the $\alpha$-order derivatives of the Gibbs free with $\alpha>4/3$ all diverge at the critical point. This means that the phase transition at the critical point is of $4/3$ order, but not usually recognized second order.

Essentially, it is the equation of state that determines the order of the phase transition at the critical point. Firstly, for the charged AdS black hole and Van der Waals fluids, one can solve the equation of state for $\nu=\nu(t,p)$ and then further expand it as series of $t$ and $p$. You will get a Puiseux series, which leads to the fractional phase transition if considering the generalized Ehrenfest classification. It is just a coincidence that the two systems both have the $4/3$-order phase transition at the critical point, because the cubic equation and the quartic equation happen to have the similar Puiseux series\cite{Sturmfels.2000}.  Secondly, near the critical point $p$ and $t$ are generally not independent due to the constraint of the equation of state. This will affect the order of the phase transition. For the charged AdS black hole and Van der Waals fluids, it is approximately a linear relation $p\approx kt$. One can also consider rotating AdS black hole, higher-dimensional black hole and black holes in modified gravities\cite{Gunasekaran.2012,Altamirano.2013,Frassino.2014,Hennigar.2017}, which have more complicated equation of state and more richer phase structures. We expect that they may supply more information on the fractional phase transition.

\acknowledgments
We would like to thank Prof. Hilfer for useful correspondence. We also thank Prof. Ren Zhao for useful discussion. This work was supported by the National Natural Science Foundation of China (Grants No. 11605107).


%
%

\appendix

\section{Introduction to the fractional derivatives}\label{fd}

The fractional derivative of a general function $f(x)$ can also be defined, but before that we first introduce some notations. We define
\be
D\equiv \frac{d}{dx} ~~\text{and}~~~ _aD_x^{-1}f(x)\equiv (I_af)(x)= \int_a^x f(t)dt,
\ee
where $a$ is called the reference point or fiducial point. 
Obviously, we have $D \cdot {}_aD_x^{-1} f(x)=f(x)$, but
$_aD_x^{-1}\cdot Df(x)=f(x)-f(a)$. Therefore, the orders of the operators are important in the calculation of derivatives.

From the  first-order derivative
\be
Df(x) \equiv \lim_{h\rightarrow 0}\frac{f(x)-f(x-h)}{h}, \no
\ee
one can obtain the higher-order derivatives
\be
D^mf(x)=\lim_{h\rightarrow 0}\frac{1}{h^m}\sum_{k=0}^{m}(-1)^k C_m^k f(x-kh),\no
\ee
or write it in another way
\be\label{mD}
{}_aD_x^mf(x)=\lim_{N\rightarrow \infty}\left[\frac{N}{x-a}\right]^m\sum_{k=0}^{N-1}(-1)^k C_m^k f(x-k\frac{x-a}{N}),
\ee
where $h=(x-a)/N$ and $C_m^k=\frac{m!}{k!(m-k)!}=\frac{\Gamma(m+1)}{\Gamma(k+1)\Gamma(m-k+1)}$. One can relax the requirement that $m$ be an integer and generalize Eq.(\ref{mD}) to
\be\label{gruI}
{}_aD_x^{-\mu}f(x)\equiv\lim_{N\rightarrow \infty}\left[\frac{x-a}{N}\right]^\mu\frac{1}{\Gamma(\mu)}\sum_{k=0}^{N-1}(-1)^k \frac{\Gamma(k+\mu)}{\Gamma(k+1)} f(x-k\frac{x-a}{N}),~~~\mu>0.
\ee
This is called Gr\"{u}nwald definition of fractional integral. It can be proved that this formula is equivalent to a simple integral
\be\label{RLI}
{}_aD_x^{-\mu}f(x) \equiv (I_a^\mu f)(x)= \frac{1}{\Gamma(\mu)}\int_a^x(x-y)^{\mu-1}f(y)dy, ~~~\mu>0,
\ee
which is called Riemann-Liouville integral. In fact, when $\mu>0$,  there are the left-handed and right-handed Riemann-Liouville integral mathematically, which are\cite{Samko.1993,Hilfer.2000}
\bea
&&(I_{a^{+}}^{\mu}f)(x) \equiv \frac{1}{\Gamma(\mu)}\int_a^x(x-t)^{\mu-1}f(t)dt, ~~~~x>a, \no \\
&&(I_{a^{-}}^{\mu}f)(x) \equiv \frac{1}{\Gamma(\mu)}\int_x^a(t-x)^{\mu-1}f(t)dt, ~~~~x<a.
\eea


On the basis of the fractional integral, the (left-handed and right-handed) Riemann-Liouville fractional derivative of $\alpha$ order can be directly defined,
\be\label{RLD}
\left.D_x^{\alpha}f(x)\right|_{RL}\equiv D^n\cdot D_x^{-(n-\alpha)}f(x)= D^n\cdot I^{n-\alpha}f(x),~~~(n-1<\alpha\leq n),
\ee
where $n$ is an integer. In this paper we always set the reference point $a=0$. So we will drop the subscript $a$ for clarity. When $\alpha$ is an integer, the above fractional derivative return back to the usual integer-order derivative.
For the Riemann-Liouville fractional derivative, there is an important property:
\be
\left.D^{\alpha}  t^{\gamma}\right|_{RL}=\frac{\Gamma(\gamma+1)}{\Gamma(\gamma+1-\alpha)} t^{\gamma-\alpha},~~~ \text{for}~\alpha>0,~\gamma>-1,~t>0.
\ee

Another commonly used definition of fractional derivative was introduced by Caputo\cite{Caputo.1967}, which is defined as
\be\label{caputo}
\left.D_x^{\alpha}f(x)\right|_{C} \equiv D_x^{-(n-\alpha)}\cdot D^n f(x)=I^{n-\alpha}\cdot D^nf(x), ~~~(n-1<\alpha\leq n).
\ee
Obviously, the two definitions are different. However, there is a relation between them
\be
\left.D^\alpha f(t)\right|_{RL}=\left.D^\alpha f(t)\right|_{C}+\sum_{k=0}^{n-1}\frac{t^{k-\alpha}}{\Gamma(k-\alpha+1)}f^{(k)}(0^{+}).
\ee

According to Caputo's definition, the fractional derivative of a constant is zero, namely $\left.D_x^{\alpha}1\right|_{C}=0, ~(\alpha>0)$. Moreover, In contrast to the Riemann-Liouville fractional derivative, the advantage of Caputo's definition is that when solving differential equations it is not necessary to define the fractional order initial conditions.  Below we will choose the Caputo's definition of the fractional derivative and drop the subscript ``~$C$~" at that point.

\bibliographystyle{JHEP}

\end{document}